\documentclass[twocolumn,pre,showpacs,preprintnumbers,amsmath,amssymb,floatfix,superscriptaddress]{revtex4}
\usepackage{graphicx}
\usepackage{amsmath}
\usepackage{amssymb}
\usepackage{color}

\begin{document}

\title{Tunneling and percolation transport regimes in segregated composites}

\author{B. Nigro}\email{biagio.nigro@epfl.ch}\affiliation{LPM, Ecole Polytechnique F\'ed\'erale de Lausanne,
Station 17, CP-1015 Lausanne, Switzerland}
\author{C. Grimaldi}\email{claudio.grimaldi@epfl.ch}\affiliation{LPM, Ecole Polytechnique F\'ed\'erale de
Lausanne, Station 17, CP-1015 Lausanne, Switzerland}
\author{P. Ryser}\affiliation{LPM, Ecole Polytechnique F\'ed\'erale de
Lausanne, Station 17, CP-1015 Lausanne, Switzerland}

\begin{abstract}
We consider the problem of electron transport in segregated conductor-insulator composites in which
the conducting particles are connected to all others via tunneling conductances thus forming
a global tunneling  connected resistor network. Segregation is induced by the presence of large
insulating particles which forbid the much smaller conducting fillers to occupy uniformly the three
dimensional volume of the composite. By considering both colloidal-like and granular-like dispersions
of the conducting phase, modeled respectively by dispersions in the continuum and in the lattice, we evaluate
by Monte Carlo simulations the effect of segregation on the composite conductivity $\sigma$, and show that
an effective medium theory applied to the tunneling network reproduces accurately the Monte Carlo results.
The theory clarifies that the main effect of segregation in the continuum is that of reducing the mean
inter-particle distances, leading to a strong enhancement of the conductivity. In the lattice segregation case
the conductivity enhancement is instead given by the lowering of the percolation thresholds for first and
beyond-first nearest neighbors. Our results generalize to segregated composites the tunneling-based
description of both the percolation and hopping regimes introduced previously for homogeneous disordered systems.
\end{abstract}
\maketitle

\section{Introduction}
\label{intro}

Transport properties of conductor-insulator composites are strongly influenced by the microstructural
characteristics of the composite itself. In particular, the dc conductivity $\sigma$ depends on
the volume fraction $\phi$ of the conducting constituents, on their size \cite{Jing2000} and
shape \cite{shape} as well as on their spatial arrangement in the composite \cite{Nan2010,McLachlan2007}.
All those aspects influence $\sigma$ through their effects on the  electrical connectivity of the
conductive phase, and can thus be exploited to meet specific criteria for the transport properties in
composites.

In compacted mixtures of micrometric conducting and insulating
powders \cite{Malliaris1971,Wu1997,Thommerel2002,Chiteme2003} the electrical
connectivity is established by direct contact connections between the conducting particles \cite{McLachlan2007}.
In this situation, $\sigma$ displays an insulator-conductor transition when the concentration $\phi$ of the
conducting phase is such that a macroscopic cluster of connected particles spans the entire sample,
allowing the charge carriers to flow between the electrodes.
Percolation theory \cite{Stauffer1994,Sahimi2003} describes such transition by mapping
the inter-particle electrical connections to a random resistor network, where the elemental conductances
$g$ are either $0$ when there is no contact or $g\neq 0$ when two conducting particles touch each other.
In this way, percolation theory predicts a power-law behavior of the form
$\sigma \simeq (\phi - \phi_c)^t$ for $\phi\gtrsim\phi_c$, where  $\phi_c$
is the critical volume fraction beyond which a spanning cluster is formed and $t$ is a universal transport exponent taking the value $t \simeq 2 $ for all three-dimensional systems \cite{Sahimi2003}.
For this kind of composites, $\sigma$ is thus mainly controlled by the value of $\phi_c$, which depends
upon the shape and the dispersion of the conducting particles.

In nanocomposites made of colloidal dispersions of nanometric conducting particles in an insulating
continuous matrix the predominant mechanism of transport is not
by direct contact, but rather through indirect electrical connections between particles established by quantum
tunneling \cite{Balberg2009}. For temperatures sufficiently high to neglect particle charging
effects, energetic disorder, and Coulomb interaction, the inter-particle conductance decays
exponentially for distances between particles larger than a characteristic tunneling
length $\xi$, which measures the electron wave function decay
within the insulating phase. Although $\xi$ depends on specific material properties,
its value is nevertheless limited to a few nanometers or less, which is a relevant
length scale for composites with nanometric conducting particles, whose typical sizes range from tens up to hundreds of nanometers.
 Hence, besides the effect
of shape and dispersion of the conducting constituents, the composite conductivity $\sigma$ is
in this case affected also by the mean distance between the particles, which essentially depends on
$\phi$.

It is clear that tunneling conduction, albeit decaying fast, does not imply a sharp cutoff of the connectivity
between particles and the introduction of a contact-like connectivity criterion, as done for powder
mixtures, is not suitable \cite{Ambrosetti2010a}. Conversely, inter-particle conductance by contact can be seen as a limiting
case of tunneling when the particle sizes $D$ are much larger than the tunneling decay length $\xi$,
as it is the case for mixtures of micrometric conducting and insulating powders \cite{Wu1997,Thommerel2002,Chiteme2003}.
It turns out that it is indeed possible to describe both contact and distance dependent connectivity
mechanisms within a single formalism, in which the conducting particles are all electrically
connected to each other by tunneling \cite{Ambrosetti2010b}. By using this global tunneling network (GTN) approach,
percolation properties of compacted powders or of other granular materials with micrometric
conducting grain sizes can then be recovered by the $D/\xi\gg 1$ limit of the theory, while
hopping transport of colloidal nanocomposites is obtained by much smaller $D/\xi$ values.

As noticed in Ref.~\cite{Ambrosetti2010b}, $D/\xi$ is not, however, the only factor discriminating
between percolation and hopping regimes. Indeed, for $D/\xi$ sufficiently large, the composite
microstructure plays the most relevant role and, through the arrangement of the
conducting matter in the composite, promotes one regime or the other. For example, homogeneous
dispersions of impenetrable conducting spheres in the (insulating) continuum are always characterized
by an hopping (or, equivalently, tunneling) type of transport. In this case indeed the conductivity
decreases fast but continuously as $\phi$ is reduced because, in average, the inter-sphere distances
increase. Instead, percolation-like behavior of transport arises in close-packed mixtures of conducting and
insulating spheres because contact or near-contact clusters of conducting spheres span the entire system
for all volume fractions larger than $\phi_c$. For large $D/\xi$, multiple percolation thresholds due
to clusters of further next-nearest neighbors can arise in fractionally occupied periodic lattices.

In this article we extend the study of the percolation and hopping regimes to the case of (locally)
non-homogeneous dispersions of conducting particles by considering segregation of the conducting
fillers due to large (compared to the conducting particles size) insulating
inclusions \cite{Kusy1977,He2004,Johner2009,Nigro2011}. This type of microstructure is rather common
in real composites and is at the origin of the large conductivity values measured also for very
low contents of the conducting phase. We shall show how this peculiar behavior arises
in both hopping and percolating regimes by solving numerically the tunneling network equations
for continuum and lattice segregated particle distributions. Furthermore, by using a generalization
\cite{Ambrosetti2010b,Grimaldi2011} of the classical effective medium approximation (EMA) 
\cite{Sahimi2003,Kirkpatrick1973,EMAref}, we explicitly relate the microstructure properties
of the composites with the transport behavior, and provide an approximate but accurate analytic
treatment of the conductivity problem, thus extending our previous results for the
continuum \cite{Nigro2011} and generalizing the formulation to the segregated case.

In Sec.~\ref{continuum} we consider the continuum regime by describing the model for the
segregated composites and by introducing the EMA formulation for the calculation of the overall
conductance. In Subsec.~\ref{continuumMC}  we present the results of both EMA calculation and
MC simulations for some segregated systems and in  Subsec.~\ref{deltac} we provide an explicit
approximate analytical formula for the composite conductivity based on EMA.
Section.~\ref{lattice} is devoted to the calculation of both EMA and MC for lattice segregated
dispersions of conducting fillers. Finally, Sec.~\ref{conclusion} is devoted to the conclusions.

\section{Segregation in the continuum}
\label{continuum}
\begin{figure}[t!]
\begin{center}
\includegraphics[scale=0.5,clip=true]{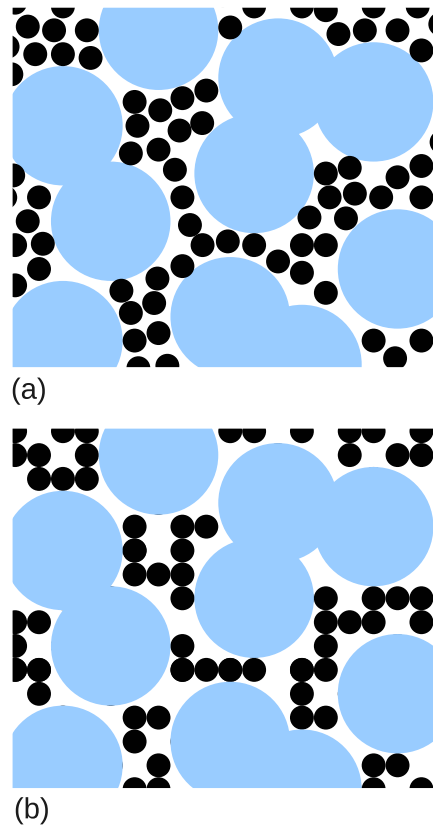}
\caption{(Color online) Schematic two-dimensional representation of the segregated dispersions in the
continuum (a) and in the lattice (b). The small black circles denote the conducting spheres of diameter $D_1$
while the larger penetrable circles are the insulating spherical inclusions of diameter $D_2$.
In (b) the centers of the conducting spheres occupy a fraction of the site of a periodic lattice. }\label{fig1}
\end{center}
\end{figure}

Segregation is very common in composite materials such as RuO$_2$-based cermets \cite{RuO2}
or polymer-based composites \cite{polymers_segre}.
This effect is particularly evident when the mean size of the insulating grains (glass-particles or
polymeric inclusions as large as few micrometers) is much larger than
that of the conducting fillers, whose typical size ranges from tens to hundreds of nanometers.
The presence of such large insulating inclusions reduces the volume available for the conducting
fillers, which are confined in the remaining space, leading thus to a locally non-homogeneous
distribution of the conducting phase in the composite.

In the following we represent the conducting phase of a conductor-insulator composite by
generating dispersions of $N_1$ hard-sphere particles of diameter $D_1$ and volume
fraction $\phi_1= \rho_1 v_1$, where $v_1= \pi D_1^3 / 6$ is the volume of a single sphere,
$\rho_1 = N_1/V$ is the number density, and $V$ is the total volume.
The non-segregated (or homogeneous case) is obtained through a random sequential addition (RSA) of hard spheres
into a cubic box of side $L$, followed by an equilibration Monte Carlo (MC) process as described
in Ref.~\cite{Nigro2011}. When the RSA limit $\phi_1^{\rm max} \sim 0.38$ is
reached\cite{Sherwood1997}, the equilibrium configuration is obtained by placing initially the hard
spheres into a cubic lattice and then by relaxing the system through MC runs.

The segregated regime is schematically shown in Fig.~\ref{fig1}(a) and is composed by a mixed system
of mutually impenetrable conducting and insulating spherical particles. This is obtained by considering
a random dispersion of $N_2$ fully penetrable spheres of diameter $D_2$ representing the insulating
inclusions \cite{Johner2009,Nigro2011}.
We fix the diameter ratio of the two particle species at $D_2/D_1=8$,
having noticed \cite{Nigro2011} that this condition is sufficient to characterize segregated systems in the
$D_2 \gg D_1$ regime. In addition, in order to minimize size effects,  $L$ is chosen as to be at
least one order of magnitude larger than $D_2$.
Since the insulating inclusions are placed randomly and are penetrable, 
the  occupied volume fraction is $\phi_2 = 1-\exp (-v_2 \rho_2)$ \cite{Torquato2002},
where $v_2 = \pi D_2^3 /6 $ and $\rho_2 = N_2/L^3$.

After having placed the insulating spheres, the remaining available space is filled with
given densities $\rho_1$ of conducting hard spheres by means of the same placement and equilibration
procedures as for the homogeneous case.
Given the mutual impenetrability of the two kinds of spheres, the available volume fraction for arranging the centers
of the conducting fillers is given by $\upsilon^*=\exp(-V_{\rm ex} \rho_2)$, where $V_{\rm ex}= \pi (D_1+D_2)^3/6$
is the excluded volume of an insulating sphere with respect to a conducting one. By exploiting the previous
definition of $\phi_2$, $\upsilon^*$ thus reduces to \cite{Johner2009}:
\begin{equation}
\label{eq:phiav}
\upsilon^* = (1-\phi_2)^{(1+D_1/D_2)^3},
\end{equation}
which defines an effective filler volume fraction $\phi_{\rm eff} = \phi_1/\upsilon^*$.

As stated in the introduction, we consider all conducting particles as electrically
connected to all others through tunneling conductances which, for two generic impenetrable
spheres $i$ and $j$ of diameter $D_1$ placed at positions $\mathbf{r}_i$ and $\mathbf{r}_j$,
is given by:
\begin{equation}
\label{eq:tunnel}
g_{ij} = g_0 \exp\!\left(-\frac{2 (r_{ij}-D_1)}{\xi}\right),
\end{equation}
where $g_0$ is a constant ``contact'' conductance which will be set equal to the unity in the following,
$\xi$ is the characteristic tunneling decay length, and $r_{ij}= |\mathbf{r}_i -\mathbf{r}_j|$ is the
distance between two conducting sphere centers.

Contrary to classical resistor networks with few connected nearest neighbors \cite{Kirkpatrick1973},
the ensemble of all tunneling conductances forms a fully connected weighted network of $N_1$
nodes, each having $N_1 -1$ neighbors. The overall conductivity depends on $D_1$ and
$\xi$, as well as on the volume fraction $\phi_1$ of the particular conducting sphere distribution.
Note that the aforementioned mapping between the sphere system and the resistor
network holds for both the homogeneous and segregated dispersions of the conducting fillers
since the information about the specific distribution is implicit in the weighted links
[\textit{i.e.}, the conductances $g_{ij}$ of Eq.~\eqref{eq:tunnel}] over all sphere centers.

\subsection{EMA and MC}
\label{continuumMC}

All these dependencies, and in particular the relationship between the spatial particle
arrangements and the global transport properties, can be made explicit by employing
the EMA formulation developed in Refs.~\cite{Ambrosetti2010b,Grimaldi2011}, which is
a generalization to complete tunneling resistor networks of the classical EMA
approach \cite{Sahimi2003,Kirkpatrick1973,EMAref}.
The original tunneling network is thus replaced by an effective one where all bond conductances
are equal to $\bar{g}$, whose value is found by requiring that the effective network has the
same average resistance as the original. By considering only two-site clusters \cite{Grimaldi2011},
the following equation for $\bar{g}$ is found:
\begin{equation}
\label{eq:EMA0}
\left\langle\sum_{i\neq j}\frac{ g_{ij}-\bar{g} }{ g_{ij}+[(N_1-1)/2-1]\bar{g}}\right\rangle=0.
\end{equation}
where $\langle\ldots\rangle$ indicates a configurational average. By introducing the radial
distribution function (RDF) $g_2(r)$ of the $N_1$ conducting hard spheres defined as \cite{Hansen}:
\begin{equation}
\label{rdf}
\rho g_2(r)=\frac{1}{N_1}\int\! \frac{d\Omega\ }{4\pi} \left\langle\sum_{i\neq j}
\delta({\bf r}-{\bf r}_{ij})\right\rangle,
\end{equation}
equation \eqref{eq:EMA0} can be recast as follows
\begin{equation}
\label{EMA1}
\int_0^\infty\! dr \frac{4\pi r^2\rho_1\, g_2(r)}{g^*\exp[2(r-D_1)/\xi]+1}=2,
\end{equation}
where $g^* \simeq N_1 \bar{g}/2$ is the conductance between any two nodes of the effective
network.
\begin{figure}[t!]
\begin{center}
\includegraphics[scale=0.42,clip=true]{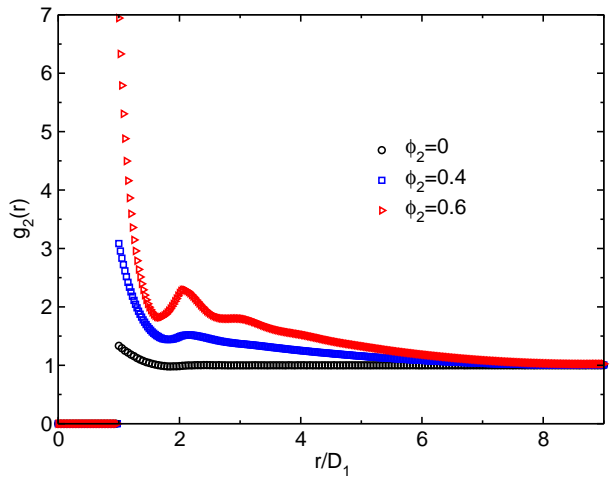}
\caption{(Color online) Numerical RDF as a function of the distance in the continuum regime
for three different volume fractions of the insulating phase $\phi_2=0$, $0.4$, and $0.6$, for $D_2/D_1=8$,
and for a fixed filler volume fraction $\phi_1=0.114$}\label{fig2}
\end{center}
\end{figure}
\begin{figure*}[t!]
\begin{center}
\includegraphics[scale=0.7,clip=true]{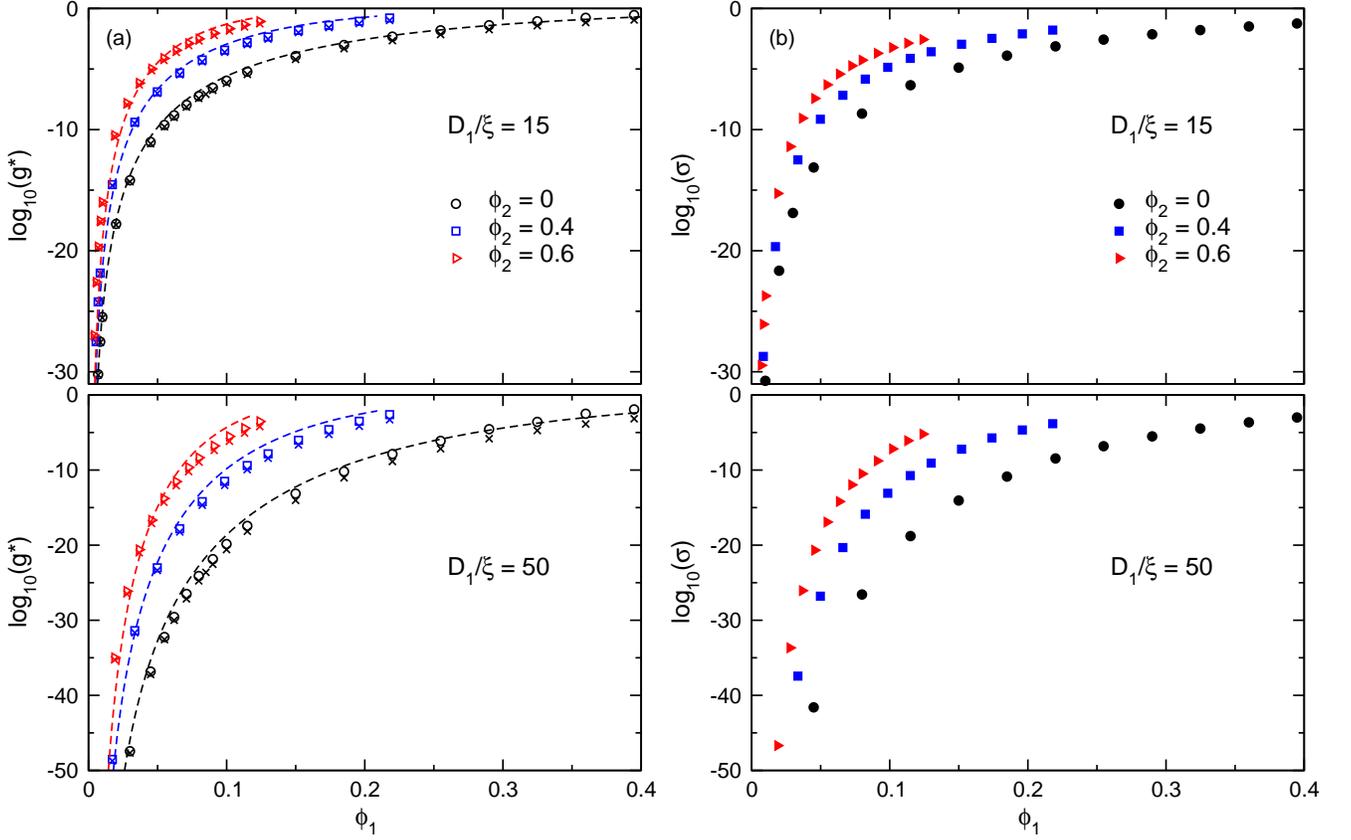}
\caption{(Color online) (a) Calculated EMA conductance $g^*$ (open symbols) as a function of the volume fraction
$\phi_1$ of the conducting spheres of diameter $D_1$ for $D_1/\xi=15$ and $50$, and for
different values of the volume fraction $\phi_2=0$, $0.4$, and $0.6$ of the insulating spheres
with diameter $D_2$ and diameter ratio $D_2/D_1=8$. The crosses and the dashed lines refer respectively
to Eq.~\eqref{gstar}, where $r^*$ is such that $Z(r^*)=2$ is satisfied, and to Eq.~\eqref{gstar_formula}.
(b): conductivity $\sigma$ for the same parameters as in (a) obtained from the numerical solution
of the tunneling resistor equations.}\label{fig3}
\end{center}
\end{figure*}
According to Eq.~\eqref{EMA1}, the whole information about the spatial distribution of the conducting
particles is contained in $g_2(r)$, which therefore directly governs the behavior of the EMA
conductance $g^*$. To illustrate how the segregation affects $g^*$ through its effects on $g_2(r)$,
we plot in Fig.~\ref{fig2} the numerically calculated RDF for $\phi_1=0.114$ and for three different
values of the volume fraction $\phi_2$ of the insulating spherical inclusions.
The $\phi_2=0$ case corresponds to a homogeneous dispersion of hard spheres in the continuum,
and the resulting RDF (circles) is basically featureless for $r>D_1$ (apart for a slight increase near
contact) because of the rather low value of $\phi_1$ used. For $\phi_2>0$ two features emerge.
First, the RDF develops an oscillating behavior with a clear peak at $r\simeq 2 D_1$ and a second one,
visible for $\phi_2=0.6$, at $r\simeq 3D_1$. Second, the RDF is enhanced with respect to the
homogeneous case for all values of $r$ lower than $r\simeq D_2=8 D_1$. These characteristics
resemble in part those observed in the RDF of the smaller particles in hard-core mixtures \cite{Yu2002}
and are indications of the enhanced probability of having particles separated by a distance $r< D_2$
when $\phi_2>0$, which means that the conducting fillers are in average closer to each other when
segregation increases. Due to the exponential dependence on the particle separation $r_{ij}$
of the tunneling conductance, Eq.~\eqref{eq:tunnel}, the EMA conductance at fixed $\phi_1$ is thus
expected to be increased by segregation because, for $\phi_2>0$, particles have lower average
value of $r_{ij}$.

This is confirmed by the results plotted in Fig.~\ref{fig3}(a), where the EMA conductance $g^*$,
obtained from solving Eq.~\eqref{EMA1} with $g_2(r)$ evaluated from MC calculations, is reported
as a function of $\phi_1$ and for $\phi_2=0$, $0.4$, and $0.6$. The role of segregation in enhancing
$g^*$ is made even more evident when $D_1/\xi$ increases, as it can be inferred by comparing
the results in the upper panel of Fig.~\ref{fig3}(a), obtained for $D_1/\xi=15$, with those with
$D_1/\xi=50$ in the lower panel.

The result that the EMA formulation provides a transparent relation between the microstructure
of segregated continuum composites, contained in $g_2(r)$, and the transport behavior is made
even more firm by the fact that the EMA conductance is in excellent accord with our fully numerical
calculations of $\sigma$ \cite{Note1}. These are shown in Fig.~\ref{fig3}(b), and have been obtained by the same
numerical procedures described in Ref.~\cite{Nigro2011} consisting in solving numerically
the Kirchoff equations of the tunneling resistor network with conductances given by Eq.~\ref{eq:tunnel}.
For the segregated cases, each symbol in Fig.~\ref{fig3}(b) is the outcome of $N_R=200$ realizations
for $N_1$ conducting particles ranging from a few hundreds for low $\phi_1$  to hundreds of thousands
for the  higher  values, by keeping $L$ fixed. Instead for the homogeneous case ($\phi_2=0$)
$N_{R}=500$ has been considered, with $N_1$ fixed at $\simeq 1000$.

\subsection{EMA analytical formula}
\label{deltac}

The good agreement between the EMA results and the fully numerical $\sigma$ in both
the homogeneous and segregated regimes suggests that further insights can be gained
directly from the EMA equation \eqref{EMA1}.
Hence we proceed here with some further approximations in the attempt to find an analytical
expression for the EMA conductance.
We start by noticing that the integral in Eq.~\eqref{EMA1} can be rewritten as
\begin{equation}
\label{EMA2}
\int_0^\infty\! dr 4\pi r^2\rho_1\, g_2(r) W(r)=2,
\end{equation}
where
\begin{equation}
\label{rstar}
W(r)=\frac{1}{\exp \left[\frac{2}{\xi}(r-r^*)\right]+1},
\end{equation}
and $r^*$ is defined by the following relation:
\begin{equation}
\label{gstar}
 g^*=\exp\left[- \frac{2}{\xi}(r^* -D_1)  \right].
\end{equation}
We note that for large values of $D_1/\xi$, which is the regime of practical interest for our
purposes \cite{Nigro2011} (see also Sec.~\ref{intro}), $W(r)$ is well approximated by
$\theta(r^*-r)$, where $\theta(x)$ is $1$
for $x\geq 0$ and
$0$ otherwise.
Thus, by adopting the definition of the cumulative coordination number
\begin{equation}
\label{zeta}
Z(r)=\int_0^r dr' 4 \pi r'^2 \rho_1 g_2(r'),
\end{equation}
which gives the number of spheres whose centers are within a distance $r$ from the center of a given sphere,
we obtain the following approximation of Eq.~\eqref{EMA1}:
\begin{equation}
\label{zr2}
 Z(r^*)=2.
\end{equation}
\begin{figure}[t!]
\begin{center}
\includegraphics[scale=0.42,clip=true]{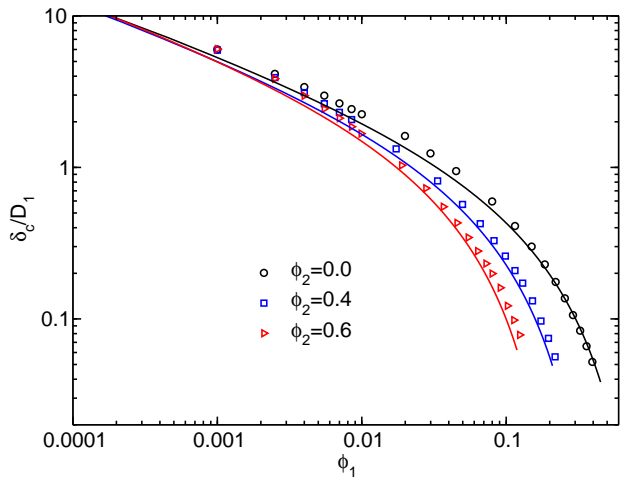}
\caption{(Color online)  Critical distance $\delta_c$ dependence on the
volume fraction $\phi_1$ of the conducting  spheres for $\phi_2=0$, $0.4$, and $0.6$ and
for $D2 /D1 = 8$ extracted from MC calculations \cite{Nigro2011}.
Solid lines: our $\delta^*$ approximation from Eq.~\eqref{rstar_app} ($\phi_2=0$) and from
Eq.~\eqref{scaling} ($\phi_2 \neq 0$).}\label{fig4}
\end{center}
\end{figure}
Equation~\eqref{gstar}, where $r^*$ is such that Eq.~\eqref{zr2} is satisfied, is plotted in Fig.~\ref{fig3}(a)
(star symbols) and is in close agreement with the numerical solution of Eq.~\eqref{EMA1}.
We proceed further by noticing that the quantity $\delta^* \equiv r^* -D_1$ is the EMA equivalent
of the critical distance $\delta_c$  found by the critical path approximation (CPA) for the tunneling
conductivity \cite{Ambegaokar1971,Strelniker2005}.
The CPA $\delta_c$, plotted in Fig.~\ref{fig4} (symbols) as a function of $\phi_1$ for
$\phi_2=0$, $0.4$ and $0.6$, is the shortest among the interparticle distances such that the subnetwork
formed by those particles having $r_{ij}-D_1\leq\delta_c$ forms a percolating cluster.
Equivalently, $\delta_c$ is such that $Z(\delta_c+ D_1)=Z_c$  is satisfied, where $Z_c$ is the critical
coordination number which, for the homogeneous case $\phi_2=0$, ranges between $Z_c \simeq 2.7 $
for $\phi_1 \rightarrow 0$ and $Z_c \simeq 1.5 $ for $\phi_1 \simeq 0.5 $ \cite{Heyes2006}.
It turns out therefore that the right-hand side of Eq.~\eqref{zr2} falls well within the range of possible
$Z_c$ values, which is the ultimate reason of the good accord between the EMA approximation
of the conductance and the fully numerical $\sigma$ already noticed in Ref.~\cite{Ambrosetti2010b}
for the homogeneous case.

By noticing from Fig.~\ref{fig4} that $\delta^* \ll D_1$ for large $\phi_1$ values and that $g_2(r)=0$ for
$r<D_1$, in order to correctly capture the high density regime, we approximate the RDF in
Eq.~\eqref{zeta} with its contact value $g_2(D_1)$. In this way, Eq.~\eqref{zr2} can be rewritten as
\begin{equation}
\label{integral_app}
g_2(D_1)\int_0^{D_1+\delta^*} dr 4 \pi r^2 \rho_1 \theta(r-D_1)=2,
\end{equation}
which can be solved for $\delta^*$ leading to:
 \begin{equation}
\label{rstar_app}
\delta^*=r^*-D_1=D_1\left[1 + \frac{1}{4\phi_1 g_2(D_1)} \right]^{1/3}-D_1.
\end{equation}
By using the Carnahan-Starling formula $g_2(D_1)=(1-\phi_1/2)/(1-\phi_1)^3$ for the RDF at
contact \cite{Carnahan1969}, which is a well-known approximation used in the theory of simple liquids \cite{Hansen},
Eq.~\eqref{rstar_app} turns out to be a rather good approximation for $\delta_c$
in the whole range of densities for $\phi_2=0$
(solid line in Fig.~\ref{fig4}), and thus we can use the same approximate scaling relation that we
formulated in Ref.~\cite{Nigro2011} for the critical distance $\delta_c$ of segregated systems. Hence, if
$\delta^*(\phi_1, \upsilon^*)$ is the EMA critical distance for a segregated system parametrized by the
available volume $\upsilon^*$ of Eq.~\eqref{eq:phiav}, then
\begin{equation}
\label{scaling}
\delta^*(\phi_1, \upsilon^*) = \upsilon^{* - 1/3}\delta^*(\phi_{\rm eff}),
\end{equation}
where $\phi_{\rm eff}=\phi_1/\upsilon^*$ is the effective volume fraction for the conducting fillers
introduced in Sec.~\ref{continuum}. Equation~\eqref{scaling} then states that the EMA critical distance in the segregated
regime can be directly obtained from that of the homogeneous case calculated at $\phi_{\rm eff}$.
As shown in Fig.~\ref{fig4}, Eq.~\eqref{scaling} compares well with MC calculations of the critical distance
$\delta_c$, so that by using Eq.~\eqref{gstar} with Eqs.~\eqref{rstar_app} and \eqref{scaling} and
$g_2(D_1)$ as given by the Carnahan-Starling expression we obtain the following approximated
analytical formula of the EMA conductance:
\begin{equation}
\label{gstar_formula}
g^*=\exp \left \{-\frac{2D_1}{\xi\upsilon^{*\frac{1}{3}}}\left[
\left(\frac{1+\phi_{\rm eff}+\phi_{\rm eff}^2-\phi_{\rm eff}^3}
{2\phi_{\rm eff}(2-\phi_{\rm eff})}\right)^{\frac{1}{3}}-1     \right]\right\}.
\end{equation}
The above expression is plotted in Fig.~\ref{fig2}(a) by dashed lines and turns out to be in very
good agreement with the full solution of the EMA integral of Eq.~\eqref{EMA1}. Hence, despite of
its simplicity, Eq. ~\eqref{gstar_formula} effectively captures the conductance
behavior of both the homogeneous and segregated cases in the whole range of $\phi_1$ values,
thus generalizing the results of Refs.~\cite{Ambrosetti2010a,Ambrosetti2010b,Nigro2011}.

\section{Segregation in the lattice}
\label{lattice}

\begin{figure*}[t!]
\begin{center}
\includegraphics[scale=0.7,clip=true]{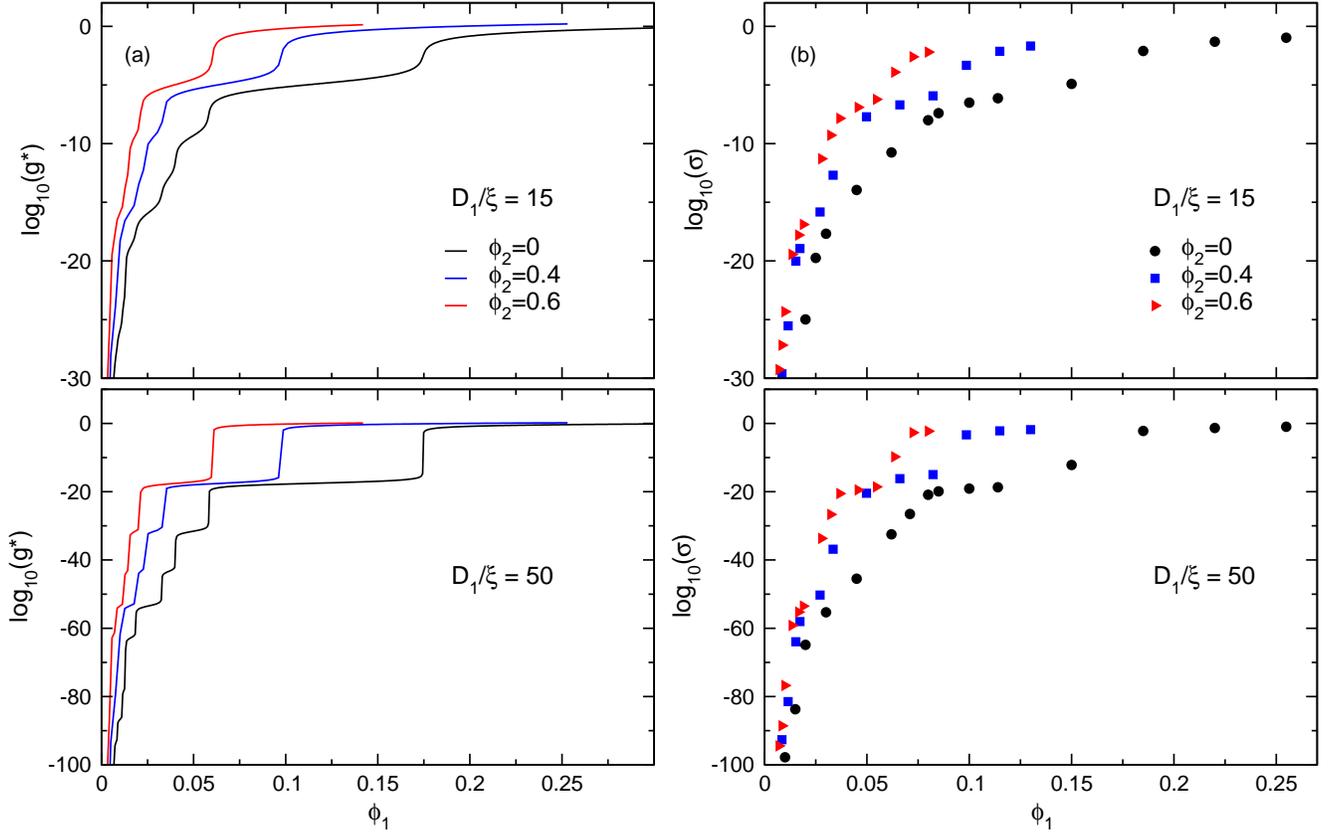}
\caption{(Color online) (a) Calculated EMA conductance $g^*$ for the segregated lattice as a function of the volume fraction
$\phi_1$ of the conducting spheres for $D_1/\xi=15$ and $50$, and for different values of the volume
fraction $\phi_2$ of the insulating spheres with diameter $D_2$. (b): conductivity $\sigma$ for the same parameters as in
(a) obtained from the numerical solution of the tunneling resistor equations.}\label{fig5}
\end{center}
\end{figure*}

Segregation can be induced in powder mixtures of conducting and insulating particles when
the mean size of the insulating grains is much larger than that of the conducting ones.
Typical examples of segregated powder mixtures and of the corresponding conductivity measurements
can be found in Refs.~\cite{Malliaris1971,Chiteme2003}. Compared to powder mixtures
of conducting and insulating grains with comparable mean sizes \cite{Wu1997,Thommerel2002}, the conductivity
of segregated powders drops by several orders of magnitude at much lower values of the volume fraction.

In order to describe the effect of segregation in powder mixtures, we consider a model composite where
the conducting particles occupy only a fraction $p$ of the total $M$ sites of a simple cubic lattice.
For non-segregated systems, we consider equal sized conducting and insulating particles of spherical shape
with diameter equal to the lattice spacing, as in the Scher and Zallen model \cite{Scher1970}. For a random distribution of the
conducting spheres on the cubic lattice the corresponding RDF reduces to \cite{Ambrosetti2010b}:
\begin{equation}
\label{rdf3}
\rho g_2(r)=\frac{p}{4\pi}\sum_{k=1,2,\ldots}\frac{\mathcal{N}_k}{R_k^2}\,\delta(r-R_k),
\end{equation}
where $\mathcal{N}_k$ is the number of the $k$th nearest neighbors being at distance $R_k$ from
a reference particle set at the origin. From Eq.~\eqref{rdf3}, and by using the tunneling inter-particle conductance
of Eq.~\eqref{eq:tunnel}, the EMA equation \eqref{EMA1} becomes
\begin{equation}
\label{EMAcube}
p\sum_{k=1,2,\ldots}\frac{\mathcal{N}_k}{g^*\exp[2(R_k-D_1)/\xi]+1}=2,
\end{equation}
whose solution is plotted in Fig.~\ref{fig5}(a) (solid black lines) as a function of the volume fraction $\phi_1$
of the conducting phase ($\phi_1=p\pi/6$) for two different values of $D_1/\xi$.
As discussed in more details in Ref.~\cite{Ambrosetti2010b}, the decrease of $g^*$ as $\phi_1\rightarrow 0$
is characterized by sharp drops at $\phi_1^k=p_k\pi/6$ where $p_k=2/(\sum_{k'=1}^k\mathcal{N}_{k'})$ is the
percolation threshold for the $k$th nearest neighbors. This behavior is due to the discrete nature of the lattice
RDF, Eq.~\eqref{rdf3}, and is confirmed by the full MC results shown in Fig.~\ref{fig5}(b),  where
the conductivity is obtained by following the same procedure of the continuum regime for
the number of particles held fixed at $N_1 \sim 1000$.

Since in real segregated powder mixtures the conducting particle sizes are in the micro-metric
range \cite{Malliaris1971,Chiteme2003}, the corresponding
large values of $D_1/\xi$ make the percolation threshold for particles at contact
$p_1=2/\mathcal{N}_1\,(=1/3\mbox{ for a cubic lattice})$ the one of practical interest.
In this regime, and for $\phi_1\gtrsim\phi_1^1$, the EMA conductance and the Monte Carlo conductivity follow the
power-law behaviors $g^*\propto (\phi_1-\phi_1^1)$ and $\sigma\propto (\phi_1-\phi_1^1)^t$, with respectively
$\phi_1^1=\pi/18\simeq 0.174$ and $\phi_1^1\simeq 0.163$, where
$t\simeq 2$ is the transport exponent for three-dimensional systems \cite{Ambrosetti2010b}.

\begin{figure}[t!]
\begin{center}
\includegraphics[scale=0.42,clip=true]{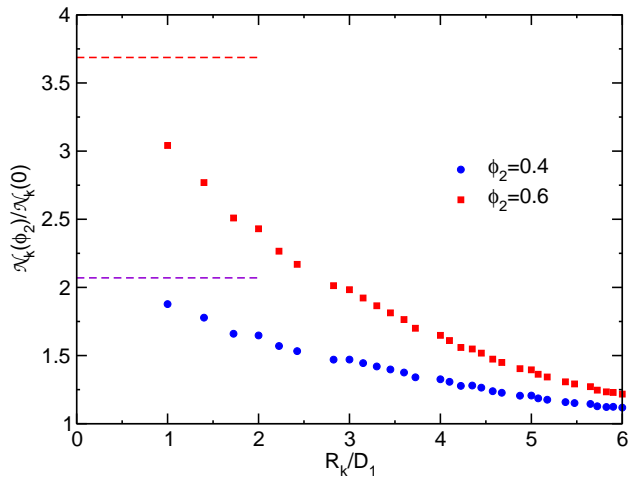}
\caption{(Color online) Enhancement ratio of the number $\mathcal{N}_k$ of $k$th neighbors for $\phi_2=0.4$ and $0.6$
(for $D_2/D_1=8$) as compared to the non-segregated case $\phi_2=0$. $R_k$ is the distance of the $k$th neighbor
lattice site from a reference site. The dashed lines are $1/\upsilon^*$}\label{fig6}
\end{center}
\end{figure}

In analogy with the continuum segregation of Sec.~\ref{continuum}, we consider the segregation in the lattice as
being due to $N_2$ penetrable and insulating spheres of diameter $D_2$ placed at random in the cubic volume.
As schematically shown in Fig.~\ref{fig1}(b),
the conducting particles will thus occupy randomly only those lattice sites lying outside the insulating spheres
and not leading to overlaps between the two species of particles. As for the homogeneous random lattice case,
the RDF of the segregated lattice is given by a series of delta-peaks centered at $R_k$, as in Eq.~\eqref{rdf3},
but with the number $\mathcal{N}_k$ of the $k$th nearest neighbors being dependent of $D_2/D_1$ and of the volume
fraction $\phi_2$ of the insulating spheres. In Fig.~\ref{fig6} we show the $R_k$ dependence of $\mathcal{N}_k(\phi_2)$
in units of the number of the $k$th nearest neighbors for the homogeneous lattice $\mathcal{N}_k(0)$ for $D_2/D_1=8$
and for $\phi_2=0.4$ and $0.6$. For $R_k\gg D_1$ the ratio $\mathcal{N}_k(\phi_2)/\mathcal{N}_k(0)$ approaches unity
which indicates that at large distances segregation plays a minor role, a result similar to the one found for the continuum
segregation case (see Fig.~\ref{fig2}). On the contrary, for $R_k$ values close to contact, segregation enhances
the number of $k$th nearest neighbors for fixed $\phi_1$ because, again in analogy with the continuum case, the reduced
available volume enhances the probability of finding conducting particles at distances lower than about $D_2$.
By presuming that the enhanced local probability can be approximated by $p_1^*=p_1/\upsilon^*$, then
$\mathcal{N}_k(\phi_2)/\mathcal{N}_k(0)$ should scale as $1/\upsilon^*$ for $R_k$ sufficiently close to contact (and $D_2/D_1$
sufficiently large), which is a fair approximation for $\phi_2=0.4$ but a less satisfactory one for $\phi_2=0.6$
(horizontal dashed lines in Fig.~\ref{fig6}).

From $p_k(\phi_2)=2/[\sum_{k'=1}^k\mathcal{N}_{k'}(\phi_2)]$ and $\mathcal{N}_k(\phi_2)>\mathcal{N}_k(0)$
it follows immediately that the percolation thresholds $p_k(\phi_2)$ for the $k$th nearest neighbors
in the segregated lattice are lower than those for the homogeneously random lattice $p_k(0)$. In particular,
the percolation threshold of particles at contact is reduced by the factor $\mathcal{N}_1(0)/\mathcal{N}_1(\phi_2)$
which from Fig.~\ref{fig6} is $0.54$ and $0.33$ for $\phi_2=0.4$ and $\phi_2=0.6$, respectively. Note that
approximating $p_1^1(\phi_2)/p_1^1(0)$ with $\upsilon^*$ leads to $0.48$ for $\phi_2=0.4$ and to $0.27$ for $\phi_2=0.6$.

The systematic lowering of $p_k(\phi_2)$ as $\phi_2$ is enhanced is reflected in the $\phi_1$-dependence of the
EMA conductance $g^*$ in Fig.~\ref{fig5}(a) obtained by solving numerically Eq.~\eqref{EMAcube} with our calculated values of
$\mathcal{N}_k(\phi_2)$. The overall effect of segregation is thus the enhancement of $g^*$ with respect to
the homogeneous lattice case at $\phi_2=0$ induced by the downshift of all $p_k(\phi_2)$ values.
This behavior is confirmed by the full MC results of Fig.~\ref{fig5}(b), which have been obtained by
solving the tunneling resistor network for $N_1$ ranging from a few hundreds for $\phi_1 \sim 10^{-3}$ to
$N_1 \sim 230000$ for $\phi_1/\upsilon^* \sim 0.5$ with $L$ held fixed at $L=10D_2$.
Given the above results and discussion, the conductivity for segregated micrometric (i.e., $D_1/\xi\gg 1$)
powders just above the percolation threshold is expected thus to follow approximately
$\sigma\propto (\phi_1-\phi_1^1\upsilon^*)^t$ where $\phi_1^1$ is the critical volume fraction for
conducting particles at contact in the absence of segregation. Due to the quasi-invariance of $\phi_1^1$, according to
which $\phi_1^1\approx 0.17$ independently of the (three dimensional) lattice topology \cite{Scher1970}, this result
is expected to apply also to non-cubic segregated lattices.

\section{Conclusions}
\label{conclusion}
In this paper we have considered the dc electrical transport problem in two-phase segregated amorphous solids, where large
insulating inclusions prevent the smaller conducting particles to be dispersed homogeneously in the three dimensional
volume. By taking into account explicitly the tunneling mechanism of electron transfer between conducting particles,
we have studied the effect of segregation for both continuum and lattice models of composites. For continuum segregated
composite materials, we have shown by theory and Monte Carlo simulations that segregation basically reduces the inter-particle
distances leading to a strong enhancement of the overall tunneling conductivity. In particular we have evidence how this
enhancement can be quasi-quantitatively reproduced by using an effective medium theory applied to the tunneling resistor network,
according to which the effect of segregation on the composite microstructure is contained entirely in the radial distribution
function for the conducting particles. By using some simple geometrical considerations in combination with the effective medium
approximation we have been able to provide an explicit formula for the composite conductivity as a function of the conducting filler
content and of the degree of segregation. When applied to our model of lattice segregation, we have demonstrated how the effective
medium theory closely reproduces the Monte Carlo results for the conductivity, which can be interpreted as due to a reduction
of the available lattice sites for placing the conducting particles.

\acknowledgements
This work was supported by the Swiss National Science Foundation (Grant No. 200021-121740).

\end{document}